\documentclass[aps,prb,twocolumn,superscriptaddress,10pt]{revtex4-1}
\usepackage[utf8]{inputenc}
\usepackage{amsmath,amssymb}
\usepackage{xcolor}
\usepackage{graphicx}

\newcommand{\R}{\text{R}}
\newcommand{\SO}{\text{SO}}
\newcommand{\bR}{\mathbf{R}}

\begin{document}
\title{A diagnosis of explicit symmetry breaking in the tight-binding constructions for symmetry-protected topological systems}
\author{Xiaoyu Wang}
\affiliation{National High Magnetic Field Laboratory, Tallahassee, FL 32310}
\author{Oskar Vafek}
\affiliation{National High Magnetic Field Laboratory, Tallahassee, FL 32310}
\affiliation{Department of Physics, Florida State University, Tallahassee, Florida 32306}

\begin{abstract}
    It has been well established that for symmetry protected topological systems, the non-trivial topology prevents a real space representation using exponentially localized Wannier wavefunctions (WFs) in all directions, unless the protecting symmetry is sacrificed as an on-site transformation. This makes it challenging to determine the symmetry of various physical observables represented using such WFs. In this work, we propose a practical method for overcoming such challenges using the Kane-Mele model as a concrete example. We present a systematic procedure for diagnosing the symmetry of any observables, as well as a method for constructing symmetric operators up to arbitrary truncation accuracy.
\end{abstract}

\maketitle

\section{Introduction}
Tight-binding models are commonly used to study interacting electron systems in the presence of a periodic lattice potential. At the heart of such models are Wannier functions (WFs) --- wavefunctions that form a complete basis for representing single-electron states \cite{girvin_yang_2019,*ashcroft_2016,VanderbiltRMP}. When the WFs are exponentially localized in all directions, the tight-binding Hamitonians can be truncated such that terms below a certain energy threshold are neglected. As a result, they have much simpler structures than their ab initio counterparts which typically contain other inactive bands, and are much easier to study both analytically and numerically.

It is well-known that non-trivial topology can prevent a simple real space representation using exponentially localized WFs. For example, an isolated band (or a band composite) with a finite Chern number --- such as a Landau level --- cannot be represented by a complete and orthonormal basis comprised of exponentially localized WFs in all directions \cite{TKNN,Marzari2007}. On the other hand, topologically non-trivial isolated bands with {\em zero} Chern number are Wannier representable, but the protecting symmetry cannot be represented as a simple site-local operation. For example, in a $Z_2$ topological insulator protected by time reversal symmetry, the two valence bands can be represented using exponentially localized WFs. However, within a unit cell, the two WFs cannot be chosen to be time-reversed partners of each other \cite{fu2006,roy2009,soluyanov11}. Similarly, in the twisted bilayer graphene, if the inter-valley scattering is neglected, the low-energy continuum model is invariant under a two-fold rotation about the $z$-axis ($C_{2z}$) followed by time reversal ($T$). The $C_{2z}T$ symmetry protects the Dirac cones in the mini-Brillouin zone which have the same chirality, leading to the so-called ``fragile topology" \cite{zou2018,po2018a,po2018,Bradlyn2017}. Such a symmetry is encoded non-locally for a tight-binding construction of the active bands \cite{kang2018,koshino2018}.

The non-locality of the protecting symmetry poses practical challenges when working with tight-binding models for symmetry protected topological systems. Truncating the model not only introduces errors to the energetics, but also breaks the protecting symmetry explicitly; albeit we expect the degree of symmetry breaking to decrease as we increase the lengthscale beyond which the tight-binding model is truncated. This has led to some recent studies on ``trivilizing" the topology by adding additional electronic bands \cite{po2018}.

In addition, we are interested in understanding the symmetry of any ``local" operators when they are represented using such WFs. Examples of such operators include the kinetic energy, the local density-density interactions, and so on. Since any truncated representation inevitably breaks the protecting symmetry, it is desired to develop a systematic procedure for differentiating between intrinsically symmetric operators from those which are explicitly symmetry-breaking. Furthermore, given any operator, we need to be able to construct its symmetric counterparts up to a pre-specified accuracy. These symmetric operators can then be used for, say, variational mean-field analysis.

To formalize our discussion, for any local operator $\hat{O}$ projected onto exponentially localized WFs in all directions, we use the notation $\sigma_{\hat{O}}$ to represent the degree of symmetry breaking when the operator is truncated, and $\delta_{\hat{O}}$ as a measure of the absolute error due to truncation. These can be defined, for example, as the matrix norms of the difference between operators. On general grounds, since the WFs are exponentially localized, all the local operators will also have exponentially small matrix elements when represented using the WFs.  As a result, $\sigma_{\hat{O}}$ induced by truncating a symmetric operator should also be exponentially small with the truncation length, and comparable to $\delta_{\hat{O}}$. On the other hand, if the operator is not invariant under the protecting symmetry, then $\sigma_{\hat{O}}$ should saturate at large truncation length, while $\delta_{\hat{O}}$ falls off exponentially.

In this paper, we first present a detailed study of the above issues using the Kane-Mele model as a concrete example. Here the topological phase is characterized by a $Z_2$ index, and protected by the time reversal symmetry. This paper is organized as follows: In section II we briefly discuss the Kane-Mele model and the projection procedure for constructing WFs, following Ref.~\cite{soluyanov11}. In section III and IV, we present a detailed analysis on the symmetry properties of truncated operators, as well as constructing symmetric operators directly in the projected Hilbert space. In section V, we demonstrate the generality of our discussions by using the one-dimensional Su-Schrieffer-Heeger (SSH) model as another example.

\section{Kane-Mele Model and projection method}
The Kane-Mele Hamiltonian is defined on a two-dimensional honeycomb lattice, given by \cite{kane_mele}:
\begin{equation} \label{eq:km_lattice}
  \begin{split}
  H & = t \sum_{\langle ij \rangle} c^{\dagger}_{i} c_{j} + i \lambda_\SO \sum_{\langle\langle ij \rangle\rangle}\nu_{ij} c^{\dagger}_{i} s^z c_{j} \\
  & + i\lambda_\R \sum_{\langle ij \rangle}  c^{\dagger}_{i} \left(\mathbf{s}\times \hat{\mathbf{d}}_{ij} \right)_z c_{j} + \lambda_v \sum_{i} \xi_i c^{\dagger}_{i}c_{i}.
\end{split}
\end{equation}
For simplicity spin index is omitted. $\lambda_v$ is a staggered chemical potential that breaks the sublattice symmetry. $\lambda_{\SO}$ is the Haldane spin orbit coupling term, and $\lambda_\R$ is the Rashba spin-orbit coupling that breaks mirror symmetry with respect to $z\leftrightarrow -z$. The local Hilbert space per unit cell is spanned by four states, representing the sublattice (A,B) and spin ($\uparrow,\downarrow$) degrees of freedom. As a result there are four Bloch bands. Time reversal symmetry has a local representation given by $is_2 \otimes \sigma_0 K$, where $\{s,\sigma\}$ are Pauli matrices acting in the spin and sublattice space respectively. $K$ denotes complex conjugation.

By increasing $\lambda_\SO$ relative to the other terms, the half-filled Kane-Mele model can be tuned to a topological phase protected by time reversal symmetry, where the two valence bands are separated from the two conduction bands by a band gap. The topological phase is characterized by a $Z_2$ invariant, defined by counting the number of pairs of zeros of the Pfaffian:
\begin{equation}
    P(\mathbf{k}) = \text{Pf}\left[\langle u_i(\mathbf{k})|\Theta|u_j(\mathbf{k})\rangle\right],
\end{equation}
where $|u_i(\mathbf{k})\rangle$ is the periodic part of the Bloch eigenstate of the $i$-th occupied band.

If the interaction energy between electrons are smaller than the band gap separating the conduction and valence bands, the hybridization can be neglected to leading order approximation, and as a result one can work within the sub-Hilbert space spanned by the two valence bands.

Following Ref.~\cite{soluyanov11}, it has been shown that a tight-binding description is possible even for the topological phase with an odd $Z_2$ invariant. The projection method \cite{marzari1997,VanderbiltRMP} for obtaining exponentially localized WFs starts by choosing two initial trial states localized within a unit cell. Such trial states are linear superpositions of both the conduction and valence states. Next these trial states are projected onto only the valence states, followed by an ortho-normalization procedure. If the trial states are chosen properly (i.e., the two trial states are not time reversed partner of each other \cite{fu2006,soluyanov11}), this procedure produces exponentially localized WFs. Following this procedure, the two WFs within a unit cell also do not form a time-reversed pair.

\begin{figure}
  \centering
  \includegraphics[width=\linewidth]{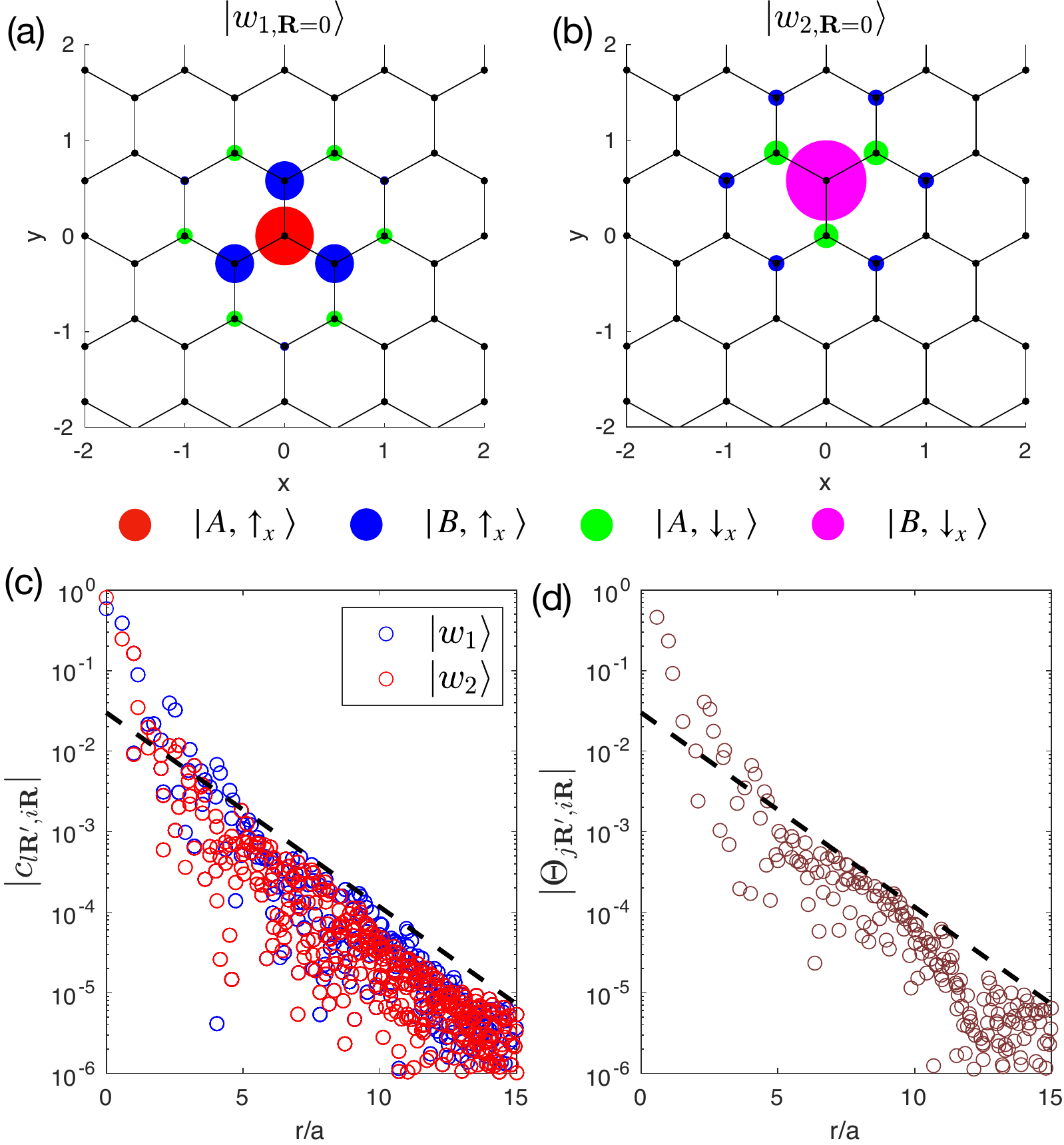}
  \caption{\label{fig:WF_localization}(a-b) The leading weights of the two WFs on honeycomb lattice. (c) The weights of the WFs plotted as a function of distance to their respective Wannier centers. (d) $|\Theta_{j\bR',i\bR}|$ plotted as a function of the relative distance. The black dashed lines in both (c) and (d) correspond to $0.03\exp\left(-\frac{r}{1.8a}\right)$, where $a$ is the lattice constant. }
\end{figure}

In Fig.~\ref{fig:WF_localization}(a-c) we discuss the properties of the WFs, following the projection method outlined above. For convenience we choose the same two trial states as Ref.\cite{soluyanov11}, namely $|A,\uparrow_x\rangle$ and $|B,\downarrow_x\rangle$, where the subscript $x$ labels the eigenstates of $s_x$ operator. The calculation is done on a $36\times 36$ periodic lattice, with parameters $t=1$, $\lambda_v=1$, $\lambda_\R=0$, and $\lambda_{\SO}=0.6$.

Fig.~\ref{fig:WF_localization}(a-b) show the weights $|c_{l\bR',i\bR}|$ of these WFs for the first few nearest neighbors to the Wannier center, defined via $|w_{i\bR}\rangle = \sum_{l=1\dots 4,\bR'} c_{l\bR',i\bR}|\chi_{l\bR'}\rangle$, where $|\chi_{l\bR'}\rangle$ denotes the four basis states of the original Hilbert space. It is clear that the WFs span more than one site. Depending on the parameter regime, the weights on the neighboring sites can be comparable or even bigger than at the Wannier center. Similar behavior have also been seen in the Wannier construction for the low-energy bands of twisted bilayer graphene, where the dominant weights of the WFs reside away from the Wannier center, having the shape of a fidget spinner\cite{kang2018,koshino2018}. At long distances, the WFs show exponential localization, as depicted in Fig.~\ref{fig:WF_localization}(c). This exponential behavior shows that it is possible to achieve a localized WF representation of the two valence bands in the topological phase.

\begin{figure*}
  \centering
  \includegraphics[width=0.9\linewidth]{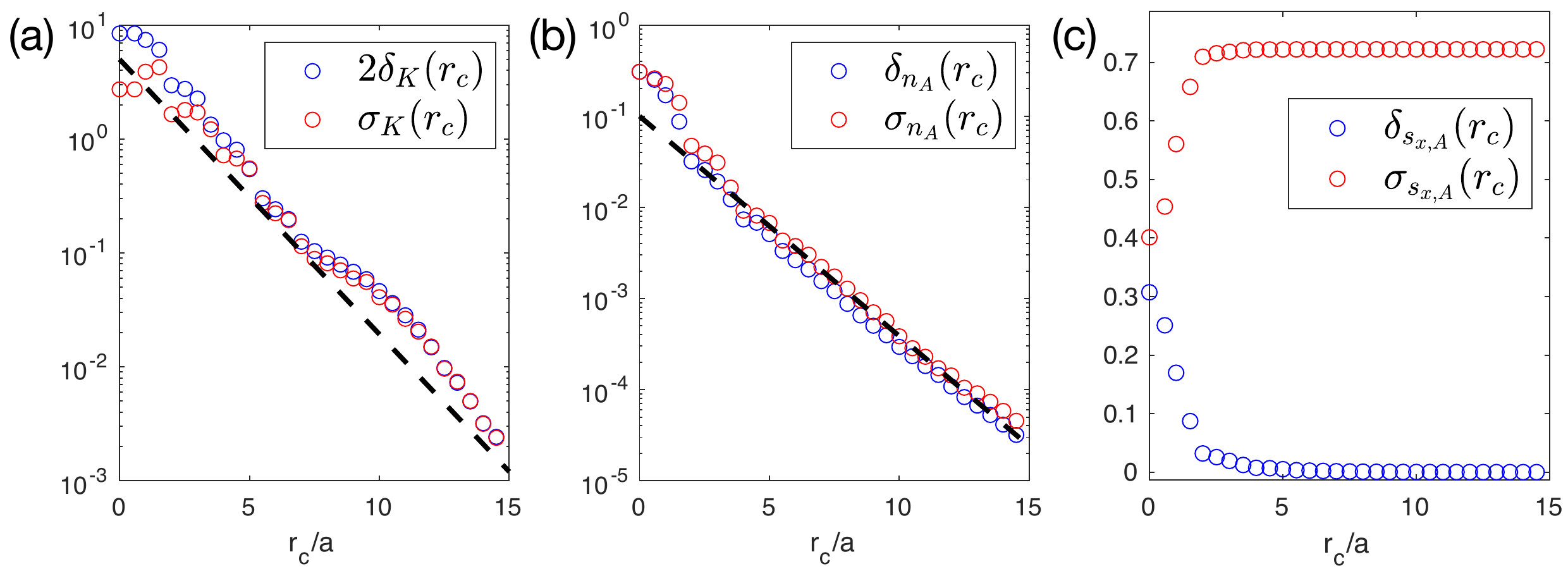}
  \caption{\label{fig:trunc_error}Truncation accuracy $\delta_\mathcal{O}(r)$ (blue circles) and degree of time reversal symmetry breaking $\sigma_\mathcal{O}(r)$ (red circles) as a function of truncation distance. (a) the projected Kane-Mele Hamiltonian $K$. Black dashed line corresponds to $5\exp\left(-\frac{r_c}{1.8a}\right)$.  (b) The total occupation number on site A of a unit cell. Black dashed line corresponds to $0.1\exp\left(-\frac{r_c}{1.8a}\right)$. (c) $s_x$ operator on site A of a unit cell.}
\end{figure*}

We proceed to discuss properties of the time reversal symmetry operator acting on these WFs. Since time reversal operation does not mix the conduction and valence Bloch states, the antiunitarity of the operator $\Theta^2=-1$ is preserved in the sub-Hilbert space spanned by the two WFs. Furthermore, time reversal operator acts non-locally in this basis, since the two WFs within a unit cell are not time-reversed partners of each other. The matrix elements can be calculated using the following relation:
\begin{equation}\label{eq:Theta_WF}
\begin{split}
   \langle w_{j\bR'}|\Theta |w_{i\bR} \rangle & = \sum_{l\bR_1,m\bR_1} c_{l\bR_1,j\bR'}^*c_{m\bR_1,i\bR} \\
   & \times \langle \chi_{l\bR_1}|is_2\otimes \sigma_0 K|\chi_{m\bR_1} \rangle.
\end{split}
\end{equation}
At long distances, the matrix elements of $\Theta$ also fall off exponentially, with the same length scale as that of the WFs. This is clearly seen in Fig.~\ref{fig:WF_localization}(d). The key message here is that despite that the time reversal symmetry is not achievable as a site-local transformation, there is still a sense of ``locality" from the long distance behavior of the matrix elements.

\section{Symmetry property of local operators \label{sec:projection}}

We are interested in understanding how various local operators are projected onto the WFs discussed above, and how truncation affects the accuracy of representation as well as the symmetry properties. Here we focus on fermionic bilinears of the particle-hole type, though generalizations to other operators (e.g. a pairing term, four-fermion term, etc.) are straightforward.

In the full Hilbert space, such a fermionic bilinear is represented as: $\hat{O} = \sum_{l\bR_1,m\bR_2}O_{l\bR_1,m\bR_2}|\chi_{l\bR_1}\rangle \langle\chi_{m\bR_2}|$, where $|\bR_1-\bR_2|$ is restricted to a few unit cells. The Kane-Mele Hamiltonian (Eq.~\ref{eq:km_lattice}) is one example of such an operator, with hopping terms up to next-nearest-neighbor sites. In addition, we are also interested in operators where both $\bR_1$ and $\bR_2$ are restricted to a few sites. For example, the local electron number and spin operators. The matrix elements of the projected operator can be calculated analogous to Eq.~\ref{eq:Theta_WF}:
\begin{equation}
\langle w_{j\bR'}|\hat{O} |w_{i\bR} \rangle  = \sum_{l\bR_1,m\bR_2} c_{l\bR_1,j\bR'}^*c_{m\bR_2,i\bR} O_{l\bR_1,m\bR_2}.
\end{equation}
The matrix elements for the time reversed operator can be calculated from:
\begin{equation}
    O^{\Theta} = \Theta O^* \Theta^\dagger.
\end{equation}
For convenience, the above symbols denote the operator representations in the projected basis. It is straightforward to see that the long-distance matrix elements of such local operators and their time reversed partners are both exponentially small, governed by the overlap of WFs far apart.

We define a Frobenius norm to characterize the absolute accuracy of a truncated operator:
\begin{equation}
    \delta_{O}(r_c) \equiv \left \lVert  O_{r_c} - O  \right \lVert_\text{F}.
\end{equation}
Here $O_{r_c;i\bR_1,j\bR_2}=O_{i\bR_1,j\bR_2}\theta(|\bR_1+\mathbf{a}_i-\bR_2-\mathbf{a}_j|-r_c)$ is the operator where matrix elements with a distance larger than $r_c$ are truncated. $\bR+\mathbf{a}_i$ labels position of the Wannier center of the $i$-th WF in the unit cell $\bR$. We see that $\delta_{O}(r_c)\propto \exp(-r_c/\xi)$, where $\xi$ is the localization length characteristic of the WFs.

To characterize the degree of time reversal symmetry breaking, we define a second Frobenius norm:
\begin{equation}
    \sigma_{O}(r_c) \equiv \left \lVert  O_{r_c} - O_{r_c}^\Theta  \right \lVert_\text{F}.
\end{equation}

The long distance behavior of $\sigma_{\hat{O}}$ can be used to differentiate between intrinsically symmetric local operators from those that are symmetry-breaking. In the former case, when the truncation length is progressively increased, the symmetry-breaking is bounded above by the absolute truncation accuracy $\sigma_{O}(r_c)\lesssim A \sigma_{O}(r_c)\propto \exp(-r_c/\xi)$, where $A$ is a non-universal constant \footnote{If additional symmetries are present, $\sigma_{O}(r_c)$ can be made arbitrarily small despite the non-onsite implementation of the protecting symmetry. One example is the inversion-symmetric Kane-Mele model in the absence of the sublattice potential. Here $\sigma_{O}(r_c)=0$ regardless of the truncation distance.}. However in the latter case it saturates to a finite constant.

In Fig.~\ref{fig:trunc_error}(a-b) we show the behavior of $\delta_{O}(r_c)$ and $\sigma_{O}(r_c)$ for the projected Kane-Mele Hamiltonian $K$ and a local electron number operator on site $A$ in unit cell $\bR$, both of which are time reversal symmetric. This is contrasted to the behavior of a local electron spin operator on site $A$ in unit cell $\bR$ (Fig.~\ref{fig:trunc_error}(c)), which is time reversal symmetry breaking.

\begin{figure}
    \centering
    \includegraphics[width=0.65\linewidth]{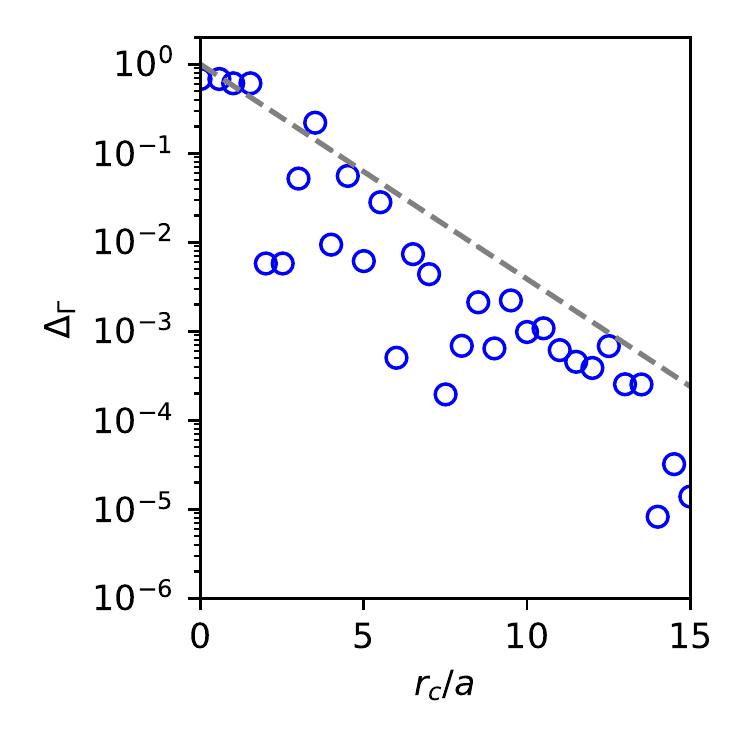}
    \caption{The single-fermion energy gap at the $\Gamma$ point $\Delta_\Gamma$ as a function of truncation length. Black dashed line corresponds to $\exp\left(-\frac{r_c}{1.8a}\right)$.}
    \label{fig:trsb_gap}
\end{figure}
Physically, the exponential smallness of $\sigma$ shows that time reversal symmetry is broken only below an exponentially small energy scale. Here this energy scale can be defined as the single-fermion energy gap at either $\Gamma$ or $M$ point of the Brillouin zone. If an exact time reversal symmetry is present, the two single-fermion states at these high symmetry points are degenerate. We illustrate the dependence of the energy gap on the truncation length in Fig.~\ref{fig:trsb_gap}.  As long as the interesting physics due to interactions occur at a higher energy scale, such explicit symmetry-breaking effects are irrelevant.

\section{Constructing symmetric operators}
\begin{figure}[t]
  \centering
  \includegraphics[width=\linewidth]{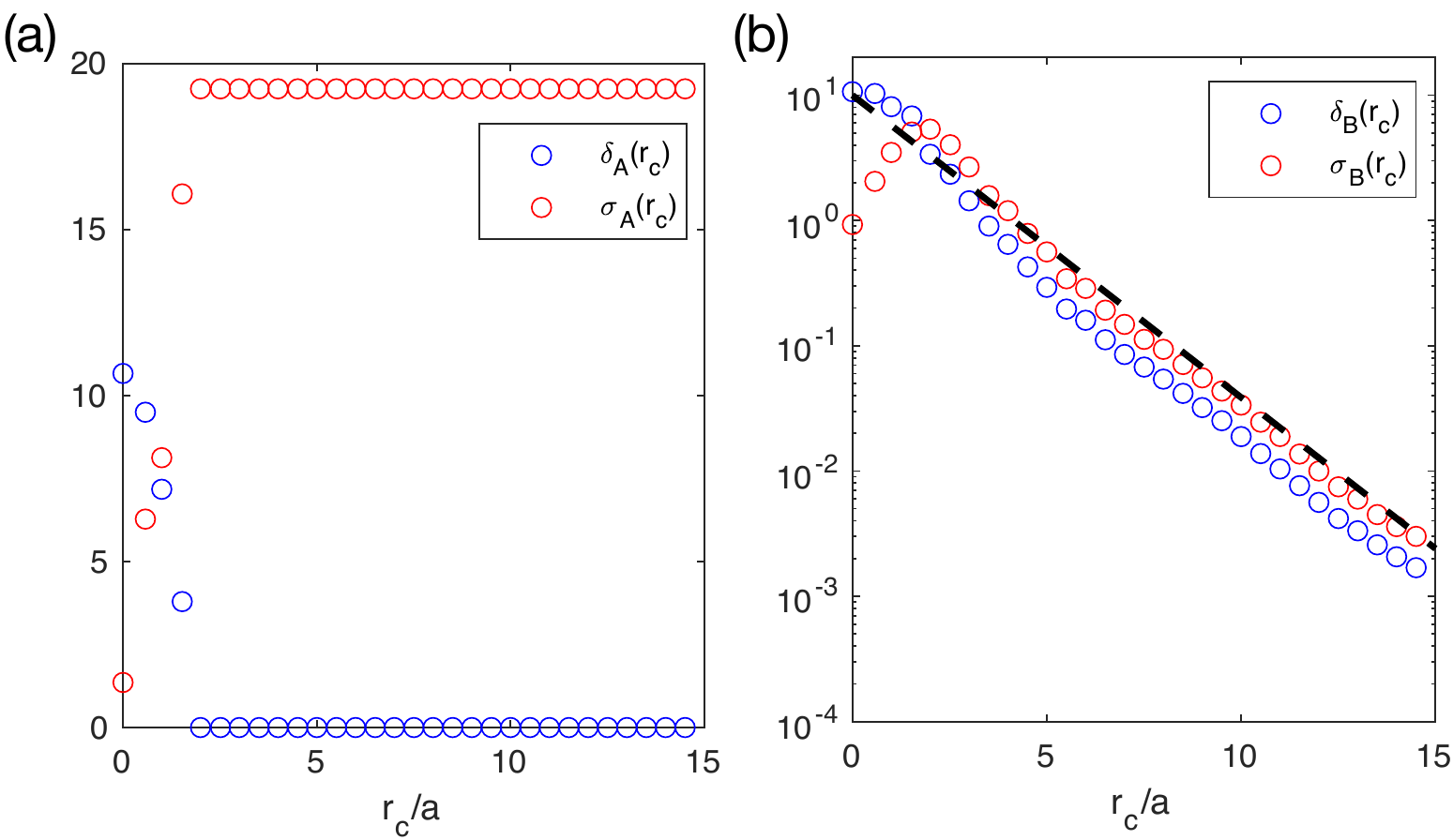}
  \caption{\label{fig:tri_trsb_diff}(a) $\delta(r)$ and $\sigma(r)$ versus truncation distance for (a) a random local operator $A$ defined up to two lattice constants, and (b) the symmetrized operator $B$. Black dashed line corresponds to $10\exp\left(-\frac{r_c}{1.8a}\right)$.}
\end{figure}

For a set of isolated bands with non-trivial topology, it is desirable to construct symmetric operators directly in the projected Hilbert space, without referencing to the full Hilbert space. These operators can be used for, say, variational mean-field analysis. Here we show that given an operator defined locally in the projected Hilbert space, the usual symmetrization procedure followed by a truncation produces a symmetric operator with exponential accuracy.

We demonstrate the above statement in the context of the Kane-Mele model. Start with an operator $A$ that is defined locally in the projected Hilbert space, we construct a time-reversal invariant operator as:
\begin{equation}
    B = \frac{1}{2}(A + \Theta A^* \Theta^\dagger).
\end{equation}
It is easy to see that $B = B^\Theta$. The symmetrization generates non-local terms that are exponentially small. As a result, both $\delta_B(r_c)$ and $\sigma_B(r_c)$ are exponentially small at long distances. This is illustrated in Fig.~\ref{fig:tri_trsb_diff} for a local operator $A$ defined with random matrix elements up to two unit cells.

\section{Wannier representation of SSH model}
In previous sections we discussed the Wannier representation of the topological bands of the Kane-Mele model, with time reversal symmetry as the protecting symmetry. However, the same physics hold generally for other symmetry-protected topological phases as well. Here as a second example, we present a similar analysis in the context of SSH model, which is one of the simplest Hamiltonians displaying non-trivial topology. The single-electron Hamiltonian is given by
\begin{align}
H &=  \frac{1}{2\pi}\int_{-\pi}^{\pi} \mathrm{d}k
    \begin{pmatrix}
    c^{\dagger}_{A,k} & c^{\dagger}_{B,k}
    \end{pmatrix}
    \mathcal{H}_k
    \begin{pmatrix}
    c_{A,k} \\ c_{B,k}
    \end{pmatrix}, \nonumber \\
\mathcal{H}(k) & \equiv \begin{pmatrix}
    0 & t_1 + t_2 e^{i ka}\\
    t_1 + t_2 e^{-i ka} & 0
    \end{pmatrix},  \label{eq:ssh}
\end{align}
where $\{A,B\}$ are two nonequivalent sites in a unit cell, $a$ is the lattice constant, and $\{t_1,t_2\}$ are real nearest-neighbor hopping parameters. The model has both a chiral symmetry ($\{\sigma_z,\mathcal{H}(k)\}=0$) and an inversion symmetry ($\sigma_x\mathcal{H}(k)\sigma_x = \mathcal{H}(-k)$).
As discussed in Ref.~\cite{neupert2018}, the SSH model describes a topological crystalline insulator protected by the inversion symmetry.  When $t_1<t_2$, it is in a topological state, characterized by a non-trivial winding number.

The low-energy state is Wannier representable. Following the procedure outlined in Ref.~\cite{soluyanov11}, it is straightforward to show that
\begin{equation}\label{eq:wannier_ssh0}
    |w_{-,R}\rangle = \frac{a}{2\pi }\int_{-\pi/a}^{\pi/a} \mathrm{d}k e^{-ikR} \frac{1}{\sqrt{2}}\left( |A,k\rangle + |B,k\rangle  e^{i\theta_k} \right),
\end{equation}
is an exponentially localized Wannier wavefunction centered at the unit cell $R$. Here $e^{i\theta_k} \equiv (t_1+t_2 e^{ika})/|t_1+t_2 e^{ika}|$. Eq.~\ref{eq:wannier_ssh0} can be rewritten as
\begin{equation}
    |w_{-,R}\rangle = \frac{1}{\sqrt{2}}\left( |A,R\rangle + \sum_{R'} c_{R',R} |B,R'\rangle  \right),
\end{equation}
where the coefficients:
\begin{equation}
c_{R',R} = \frac{1}{2\pi}\int_{-\pi}^{\pi} \mathrm{d}k e^{-ik (R-R')}e^{i\theta_k}.
\end{equation}
In the limit of $t_2\rightarrow 0$ (trivial phase), $\exp(i\theta_k)=1$, and $c_{R',R}=\delta_{R',R}$. The Wannier state describes a intra-unit-cell dimer. In the opposite limit of $t_1\rightarrow 0$ (topological phase), $\exp(i\theta_k)=\exp(ika)$, and $c_{R',R}=\delta_{R'+a,R}$. The Wannier state represents a inter-unit cell dimer. For arbitrary $t_1/t_2\neq 1$, the exponential localization of $|w_{-,R}\rangle$ is manifested in the exponential smallness of $c_{R',R}$ at large separation $|R'-R|$.
\begin{figure}
    \centering
    \includegraphics[width=\linewidth]{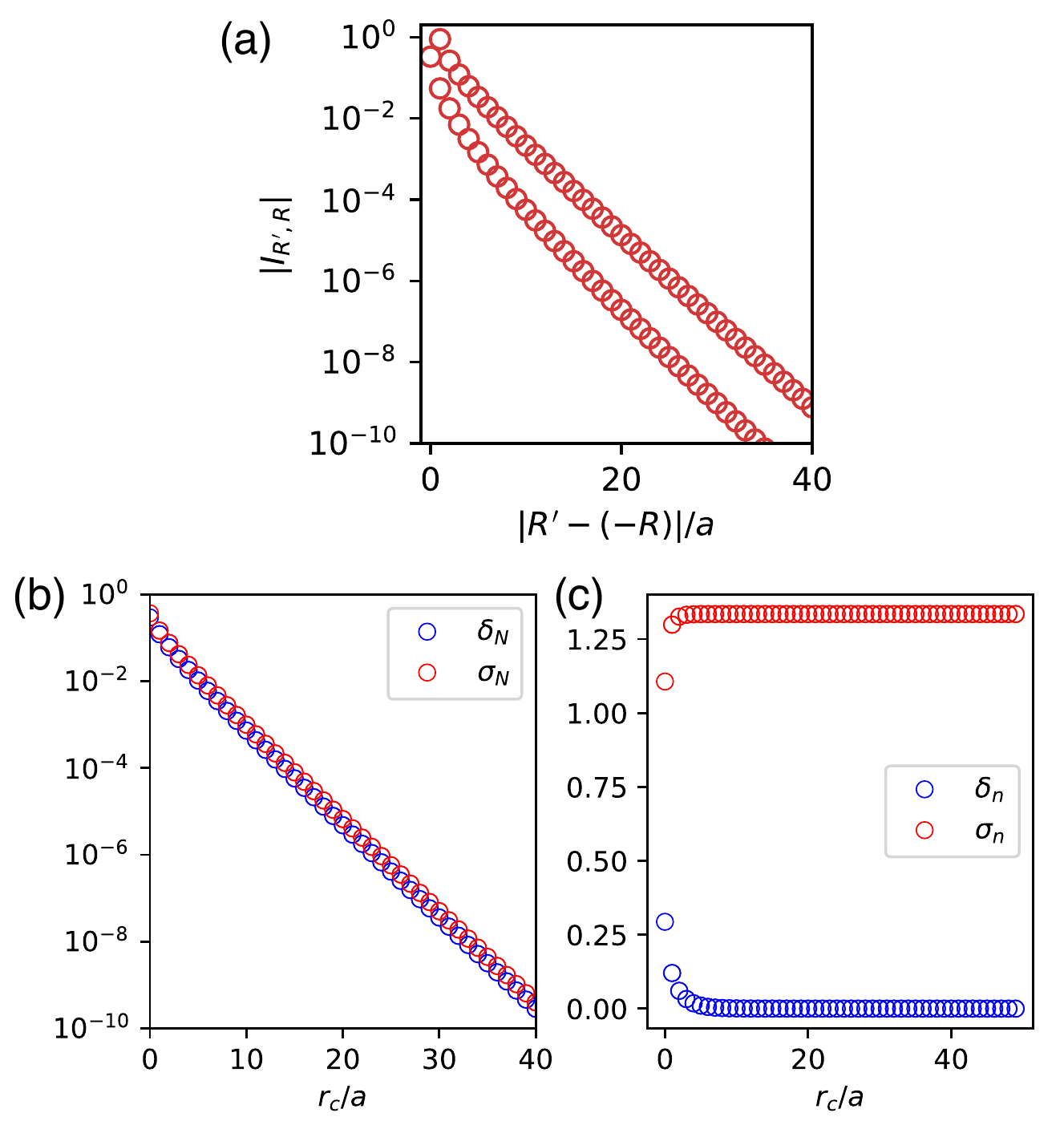}
    \caption{(a) $|I_{R',R}|$ as a function of the relative distance: $|R'-(-R)|$. (b-c) $\delta(r_c)$ and $\sigma(r_c)$ as a function of truncation length for the inverse-symmetric operator $N$ and anti-symmetric operator $n$ respectively. The numerics are performed for $t_2/t_1=1.6$.}
    \label{fig:ssh}
\end{figure}

In the Wannier basis, the inversion symmetry becomes a non-local operator. We write $I|w_{-,R}\rangle = \sum_{R'} I_{R',R}|w_{-,R'}\rangle$, and plot the coefficients as a function of separation $|R'-(-R)|$ in Fig.~\ref{fig:ssh}(a).

In Fig.~\ref{fig:ssh}(b-c), we compute $\{\delta_O(r_c), \sigma_O(r_c)\}$ for both an inversion symmetric and an antisymmetric operator. Here $\sigma_O(r_c)$ is computed according to Eq.~7, with inversion symmetry taking place of the time reversal symmetry. The inversion-symmetric operator is chosen to be $N = c^{\dagger}_{A,R=0}c_{A,R=0}+c^{\dagger}_{B,R=0}c_{B,R=0}$, and the antisymmetric operator is chosen to be $n = c^{\dagger}_{A,R=0}c_{A,R=0}-c^{\dagger}_{B,R=0}c_{B,R=0}$. It is clear that $\sigma_N(r_c)$ becomes exponentially small at large truncation length, whereas  $\sigma_n(r_c)$ saturates to a constant.

\section{Conclusion}
For a set of isolated electron bands with non-trivial topology, it is possible to achieve a real space representation using exponentially localized WFs in all directions, provided that the protecting symmetry is sacrificed as a site-local transformation. As a result, under any scheme of truncation, the representation of any local and symmetric operator using these WFs inevitably breaks the protecting symmetry explicitly. This has led to debates on the validity of tight-binding implementations for symmetry-protected topological systems \cite{po2018,po2018a,zou2018}.

Using the Kane-Mele model and the SSH model as examples, we presented a quantitative discussion of the severity of the degree of symmetry breaking. We showed that the exponential localization of the WFs guarantees that the symmetry properties for an intrinsically symmetric operator is retained with exponential accuracy. More precisely, the accuracy of symmetry is bounded by the absolute accuracy of the truncation up to a non-universal constant. As a result, a tight-binding implementation should {\em not} lead to significant issues, as long as the interesting physics occur on an energy scale larger than the exponentially small energy scale where symmetry breaking effects are important.

\section{Acknowledgement}
XW acknowledge financial support from National MagLab, which is funded by the National Science Foundation (DMR-1644779) and the state of Florida. O.V. was supported by NSF DMR-1916958.

\bibliographystyle{apsrev4-1}
\bibliography{references}

\end{document}